\documentclass[aps,prl,floats,twocolumn,showpacs,superscriptaddress,preprintnumbers,
eqsecnum,amsmath,amssymb,nofootinbib]{revtex4}

\usepackage{graphicx}
\usepackage{dcolumn}
\usepackage{bm}
\usepackage{epsfig}

\newcommand{\be}{\begin{eqnarray}}
\newcommand{\ee}{\end{eqnarray}}
\newcommand{\nn}{\nonumber}
\newcommand{\gev}{{\rm GeV}}
\newcommand{\mev}{{\rm MeV}}
\newcommand{\<}{\langle}
\renewcommand{\>}{\rangle}

\begin{document}

\preprint{
\hbox{ICCUB-09-214,~}
\hbox{RM3-TH/09-11}
}

\title{$K \to \pi \ell \nu$ Semileptonic Form Factors from Two-Flavor Lattice QCD}

\author{V.~Lubicz}
\affiliation{Dipartimento di Fisica, Universit{\`a} di Roma Tre, Via della Vasca Navale 84, I-00146 Roma, Italy}
\affiliation{Istituto Nazionale di Fisica Nucleare, Sezione di Roma Tre, Via della Vasca Navale 84, I-00146 Roma, Italy}
\author{F.~Mescia}
\affiliation{Dep.~ECM and ICC, Universitat de Barcelona, Diagonal 647, 08028 Barcelona, Spain}
\author{S.~Simula}
\affiliation{Istituto Nazionale di Fisica Nucleare, Sezione di Roma Tre, Via della Vasca Navale 84, I-00146 Roma, Italy}
\author{C.~Tarantino}
\affiliation{Dipartimento di Fisica, Universit{\`a} di Roma Tre, Via della Vasca Navale 84, I-00146 Roma, Italy}
\affiliation{Istituto Nazionale di Fisica Nucleare, Sezione di Roma Tre, Via della Vasca Navale 84, I-00146 Roma, Italy}

\collaboration{European Twisted Mass Collaboration} \noaffiliation

\begin{abstract}

We present new lattice results of the $K \to \pi \ell \nu$ semileptonic form factors 
obtained from simulations with two flavors of dynamical twisted-mass fermions, 
using pion masses as light as $260~\mev$.
Our main result is $f_+(0) = 0.9560(84)$, which, combined with the latest experimental 
data for $K_{\ell 3}$ decays, leads to $|V_{us}| = 0.2267(5)_{\mbox{exp.}}(20)_{f_+(0)}$.
Using the PDG(2008) determinations of $|V_{ud}|$ and $|V_{ub}|$ our result implies 
for the unitarity relation $|V_{ud}|^2 + |V_{us}|^2 + |V_{ub}|^2 = 
1.0004(15)$.
For the $O(p^6)$ term of the chiral expansion of $f_+(0)$ we get $\Delta f \equiv f_+(0) 
- 1 - f_2 = -0.0214(84)$.

\end{abstract}

\pacs{11.15.Ha,12.15.Hh,12.38.Gc}
\keywords{lattice QCD, CKM matrix, Kaon, semileptonic}

\maketitle

The Cabibbo's angle, or equivalently the CKM matrix element $|V_{us}|$~\cite{CKM}, 
is one of the fundamental parameters of the Standard Model.
The most precise determination of $|V_{us}|$ comes from $K \to \pi \ell \nu$ 
($K_{\ell 3}$) decay.
The PDG(2008) quotes $|V_{us}| = 0.2255(19)$ \cite{PDG}.
It is based on the new, very accurate experimental determination of the product 
$|V_{us}| f_+(0) = 0.21668(45)$ \cite{PDG,FlaviaNet} and on the old estimate 
of the vector form factor at zero-momentum transfer $f_+(0) = 0.961(8)$ 
given in Ref.~\cite{LR}.

The determination of $f_+(0)$ using lattice QCD started only recently with the quenched 
calculation of Ref.~\cite{SPQCDR}, where it was shown how $f_+(0)$ can be determined 
at the physical point with a $\simeq 1 \%$ accuracy.
The findings of Ref.~\cite{SPQCDR} triggered various unquenched calculations of 
$f_+(0)$, namely those of Refs.~\cite{JLQCD,RBC06,QCDSF} with $N_f = 2$ and pion 
masses above $\simeq 500~\mev$ and the recent one of Ref.~\cite{RBC08} with 
$N_f = 2 + 1$ and pion masses starting from $330~\mev$.

In this Letter we present a new lattice result for $f_+(0)$ obtained from simulations 
with two flavors of dynamical twisted-mass quarks, using pion masses from 
$260~\mev$ up to $580~\mev$.
Our determination of $f_+(0)$ includes the estimates of all sources of systematic 
errors: discretization, finite size effects (FSE's), $q^2$-dependence, chiral 
extrapolation and the effects of quenching the strange quark.

The chiral extrapolation and the related uncertainty are investigated using 
both SU(3) and, for the first time, SU(2) Chiral Perturbation Theory (ChPT).
Within the former one can perform a systematic expansion of $f_+(0)$ of the type 
$f_+(0) = 1 + f_2 + f_4 + ...$, where $f_n = {\cal{O}}[M_{K,\pi}^n / (4 \pi f_\pi)^n]$ 
and the first term is equal to unity due to the current conservation in the SU(3) 
limit. 
Because of the Ademollo-Gatto (AG) theorem \cite{AG}, the first correction 
$f_2$ does not receive contributions from the local operators of the effective 
theory and can be computed unambiguously in terms of the kaon and pion masses 
($M_K$ and $M_\pi$) and the pion decay constant $f_\pi$. 
It takes the value $f_2 = -0.0226$ at the physical point \cite{LR}.
The task is thus reduced to the problem of finding a prediction for 
 \be
    \Delta f \equiv f_4 + f_6 + ... = f_+(0) - (1 + f_2) ~ .
    \label{eq:deltaf}
 \ee

Recently SU(2) ChPT at the next-to-leading order (NLO) has been applied to study 
the quark-mass dependence of $f_+(0)$ \cite{SU2}.
In SU(2) ChPT the strange quark field does not satisfy chiral symmetry and the 
dependence on the strange quark mass, $m_s$, is absorbed into the low-energy 
constants (LEC's) of the effective theory.
The convergence of SU(2) ChPT is expected to be good when the $u/d$ quark mass 
is significantly smaller than $m_s$.
In the case of $f_+(0)$ one gets the NLO result \cite{SU2} 
 \be
    f_+(0) = F_+ - \frac{3}{4} \frac{M_\pi^2}{(4 \pi f_\pi)^2} \mbox{log}(\frac{M_\pi^2}{\mu^2}) 
             + c_+ M_\pi^2 + {\cal{O}}(M_\pi^4) 
    \label{eq:SU2}
 \ee
where $F_+$ and $c_+$ are LEC's functions of $m_s$ and $c_+$ depends also on the 
renormalization scale $\mu$ in such a way that the whole NLO result (\ref{eq:SU2}) 
is independent on $\mu$.

For the extrapolation of our lattice data to the physical point we apply both 
SU(2) and SU(3) ChPT obtaining consistent results, which help constraining the 
uncertainty of the chiral extrapolation.

We perform simulations with $N_f = 2$ flavors of dynamical twisted-mass quarks 
\cite{twisted_mass} generated with the tree-level Symanzik improved gauge action 
at a lattice spacing $a = 0.0883(6)~\mbox{fm}$ \cite{pion,spacing} ($\beta = 3.9$), 
for six values of the (bare) sea quark mass, namely $a m_{sea} = 0.0030, 0.0040, 
0.0064, 0.0085, 0.0100, 0.0150$ (see Ref.~\cite{ETMC}).
The valence light-quark mass is always kept equal to the sea quark mass (unitary 
pions) and the simulated pion masses goes from $\simeq 260$ to $\simeq 575~\mev$.
For each pion mass we use three values of the (bare) strange quark mass, namely 
$a m_s = 0.015, 0.022, 0.027$, to allow for a smooth, local interpolation of 
our results to the physical strange quark mass ($a m_s^{phys} \simeq 0.021$).

At the two lowest pion masses the lattice volume is $L^3 \cdot T = 32^3 \cdot 64$ 
in lattice units, while at the higher ones it is $24^3 \cdot 48$ in order to 
guarantee that $M_\pi L \gtrsim 3.7$.

We perform two additional simulations: the first one at $M_\pi \simeq 300~\mev$ 
using the smaller volume and the second at $M_\pi \simeq 470~\mev$ using a finer 
lattice spacing $a \simeq 0.07~\mbox{fm}$ ($\beta = 4.05$) in order to check 
FSE's and discretization errors, respectively.

The 2- and 3-point correlation functions relevant in this work are calculated 
using all-to-all quark propagators evaluated with the ``one-end-trick" 
stochastic procedure. 
All the necessary formulae can be easily inferred from Ref.~\cite{pion}, where 
the degenerate case of the pion form factor is illustrated in details.
At each value of the pion mass the statistical errors are evaluated with the 
jackknife procedure, while a bootstrap sampling is applied in order to 
combine the jackknives for different pion masses.

The matrix element of the weak vector current $V_\mu$ can be written as
 \be
    \< \pi(p^\prime) | V_\mu | K(p) \> & = & P_\mu ~ f_+(q^2) + q_\mu ~ f_-(q^2) ~ ,
    \label{eq:KPi}
 \ee
where $P_\mu = p_\mu + p_\mu^\prime$ and $q_\mu = p_\mu - p_\mu^\prime$, and 
the scalar form factor $f_0(q^2)$ is defined as
 \be
    f_0(q^2) & = & f_+(q^2) + \frac{q^2}{M_K^2 - M_\pi^2} f_-(q^2) ~ .
    \label{eq:f0}
 \ee

Following Ref.~\cite{SPQCDR} the scalar form factor at $q^2 = q_{max}^2 \equiv 
(M_K - M_\pi)^2$ can be calculated on the lattice with very high statistical 
precision using a suitable double ratio of 3-point correlation functions. 
In the present simulations we get a precision better than $\simeq 0.2 \%$ (see 
Table \ref{tab:fplus}).

\begin{figure}[!hbt]

\centerline{\includegraphics[scale=0.45]{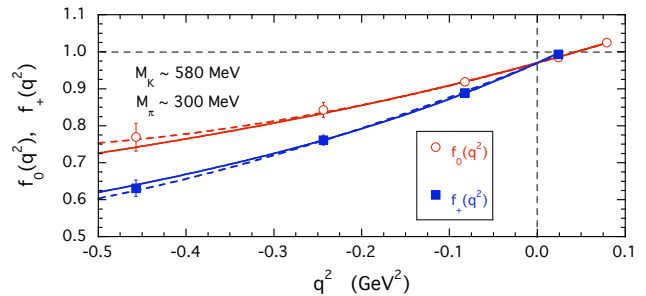}}

\caption{\it Scalar $f_0(q^2)$ and vector $f_+(q^2)$ form factors obtained at 
$M_\pi \simeq 300~\mev$ and $M_K \simeq 580~\mev$ versus $q^2$ in physical units.
The solid and dashed lines are the results of the fits based on Eqs.~(\ref{eq:pole}) 
and (\ref{eq:quadratic}), respectively.}

\label{fig:fq2}

\end{figure}

At each pion and kaon masses we determine both the vector $f_+(q^2)$ and the 
scalar $f_0(q^2)$ form factors for several values of $q^2 < q_{max}^2$ in order 
to interpolate at $q^2 = 0$.
We take advantage of the twisted boundary conditions (see Ref.~\cite{pion} for 
details) to achieve values of $q^2$ quite close to $q^2 = 0$.
The momentum dependencies of both form factors are nicely fitted either by a 
pole behavior 
 \be
    f_{+,0}(q^2) = f_+(0) / (1 - s_{+,0} ~ q^2)
    \label{eq:pole}
  \ee
or by a quadratic dependence on $q^2$
 \be
    f_{+,0}(q^2) = f_+(0) \cdot (1 + \bar{s}_{+,0} ~ q^2 + \bar{c}_{+,0} ~ q^4) ~ ,
    \label{eq:quadratic}
 \ee
where the condition $f_0(0) = f_+(0)$ is understood.
The quality of the two fits is illustrated in Fig.~\ref{fig:fq2}.

\begin{figure}[!htb]

\centerline{\includegraphics[scale=0.45]{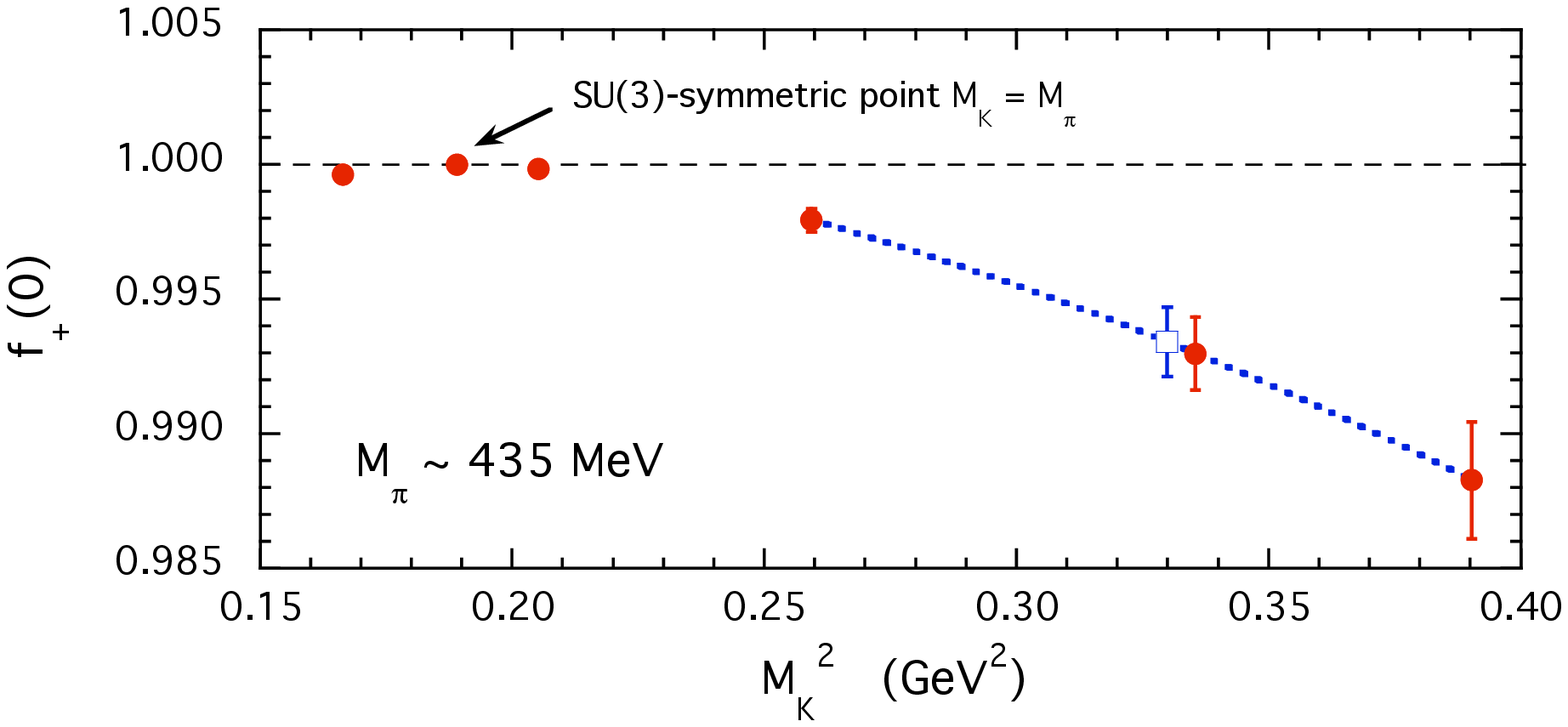}}

\caption{\it Results for $f_+(0)$ versus $M_K^2$ at $M_\pi \simeq 435~\mev$.
The square corresponds to the value of $f_+(0)$ obtained by local interpolation 
via quadratic splines (dotted line) at the reference kaon mass $M_K^{ref} 
\simeq 575~\mev$ from Eq.~(\ref{eq:MKref}).}

\label{fig:f(0)}

\end{figure}

The values obtained for $f_+(0)$ depend on both the pion and kaon masses.
The dependence on the latter is shown in Fig.~\ref{fig:f(0)} at $M_\pi \simeq 
435~\mev$ and it appears to be quite smooth, so that an interpolation at the 
physical strange quark mass can be easily performed using quadratic splines.
This is obtained by fixing the combination ($2 M_K^2 - M_\pi^2$) at its physical 
value, which at each pion mass defines a {\em reference} kaon mass $M_K^{ref}$:
 \be 
 2 [M_K^{ref}]^2 - M_\pi^2 = 2 [M_K^{phys}]^2 - [M_\pi^{phys}]^2 
 \label{eq:MKref}
 \ee
with $M_\pi^{phys} = 135.0~\mev$ and $M_K^{phys} = 494.4~\mev$.

Note that at the SU(3)-symmetric point $M_K = M_\pi$ the absolute 
normalization $f_+(0) = 1$ is imposed automatically by the double ratio method 
of Ref.~\cite{SPQCDR}.

The results for $f_+(0)$, obtained using the pole dominance (\ref{eq:pole}) or 
the quadratic fit (\ref{eq:quadratic}), and interpolated at the reference kaon 
mass (\ref{eq:MKref}), are given in Table \ref{tab:fplus} for each pion mass.
It can be seen that the values of $f_+(0)$ corresponding to different 
$q^2$-dependencies of the form factors differ by less than half of 
the statistical errors.
In what follows we will show in the figures only the results obtained using the 
pole fit (\ref{eq:pole}).

\begin{table}[!tht]

\begin{center}
\begin{tabular}{||c|c||c||c|c||}
\hline
 $M_\pi$ & $M_K^{ref}$ & $f_0(q_{\mbox{max}}^2)$ &        $f_+(0)$ &             $f_+(0)$ \\
$(\mev)$ &    $(\mev)$ &                         & $\mbox{(pole)}$ & $\mbox{(quadratic)}$ \\ \hline \hline
 $260$ &  $520$ & $1.03097~(224)$ & $0.97519~(499)$ & $0.97374~(505)$ \\ \hline
 $300$ &  $530$ & $1.01923~(121)$ & $0.98052~(440)$ & $0.97950~(390)$ \\ \hline
 $375$ &  $555$ & $1.00961~(123)$ & $0.98916~(264)$ & $0.98813~(248)$ \\ \hline
 $435$ &  $575$ & $1.00416~~(43)$ & $0.99343~(130)$ & $0.99273~(131)$ \\ \hline
 $470$ &  $590$ & $1.00272~~(34)$ & $0.99421~(79)$ & $0.99413~~(85)$ \\ \hline
 $575$ &  $635$ & $1.00016~~(~6)$ & $0.99823~~(15)$ & $0.99827~~(19)$ \\ \hline \hline

\end{tabular}

\caption{\it Results for $f_0(q_{\mbox{max}}^2)$ and $f_+(0)$, obtained with the 
pole (\ref{eq:pole}) or quadratic (\ref{eq:quadratic}) fits, interpolated at 
the reference kaon mass (\ref{eq:MKref}) for each simulated pion mass.
\label{tab:fplus}}

\end{center}

\end{table}

The SU(3) chiral analysis of $f_+(0)$ starts by considering the NLO term $f_2$, 
using the exact expression $f_2^{PQ}$ evaluated for our partially quenched 
(PQ) setup in Ref.~\cite{PQ},
 \be
   f_2^{PQ} & = & - \frac{2M_K^2 + M_\pi^2}{32\pi^2f_\pi^2} - \frac{3M_K^2M_\pi^2 
   \mbox{log}(M_\pi^2 / M_K^2)}{64\pi^2f_\pi^2 (M_K^2 - M_\pi^2)} \nn \\
   & + & \frac{M_K^2 (4M_K^2 - M_\pi^2) \mbox{log}(2 - M_\pi^2 / M_K^2)}
   {64\pi^2f_\pi^2 (M_K^2 - M_\pi^2)} ~ ,
   \label{eq:f2PQ}
 \ee
and by constructing the quantity $\Delta f$ from Eq.~(\ref{eq:deltaf}).
We then carry out the extrapolation to the physical point using a simple 
phenomenological ansatz in terms of $M_\pi^2$: 
 \be
    \Delta f = \Delta_0 + \Delta_1 M_\pi^2 + \Delta_2 M_\pi^4  + 
    \Delta_3 M_\pi^2 ~ \mbox{log}(M_\pi^2)~ ,
    \label{eq:fit}
 \ee
where $\Delta_{0,1,2,3}$ are fitting parameters.

\begin{figure}[!htb]

\centerline{\includegraphics[scale=0.45]{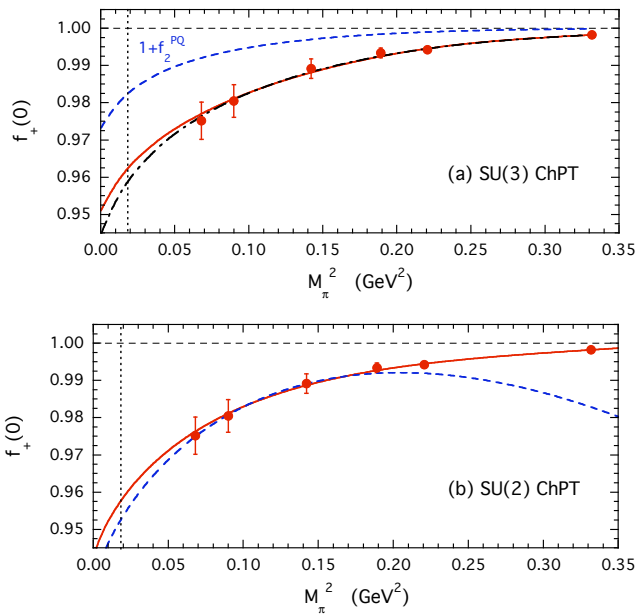}}

\caption{\it Results for $f_+(0)$ versus $M_\pi^2$ at $M_K = M_K^{ref}$ analyzed 
using SU(3) (a) and SU(2) (b) ChPT.
In (a) the SU(3) LO + NLO term, $1 + f_2^{PQ}$, is shown by the dashed line. 
The solid and dot-dashed lines are the results of the fit (\ref{eq:fit}) for 
$\Delta f$ with $\Delta_3 = 0$ and $\Delta_2 = 0$, respectively.
In (b) the dashed line is the SU(2) LO + NLO fit (\ref{eq:SU2}) applied to our 
data with $M_\pi \lesssim 0.4~\gev$, while the solid line corresponds to the 
result of fitting all lattice points adding to Eq.~(\ref{eq:SU2}) a NNLO term 
proportional to $M_\pi^4$.
The vertical line corresponds to $M_\pi^{phys} = 135.0~\mev$.}

\label{fig:fplus}

\end{figure}

The results obtained for $f_+(0)$ using two fits for $\Delta f$, one with 
$\Delta_3 = 0$ and the other with $\Delta_2 = 0$, are shown in 
Fig.~\ref{fig:fplus}(a).
It can be seen that: 
i) the (absolute) size of $\Delta f$, whose chiral expansion starts from the 
NNLO term $f_4$, is even larger than the one of the leading NLO term $f_2^{PQ}$ 
at all pion masses, and
ii) the impact of the logarithmic term at NNLO is quite small.

A relevant check on our fits (\ref{eq:fit}) is that they turn out to be consistent 
with zero (within the statistical errors) at the point $M_\pi = M_K^{ref}$, as 
required by the AG theorem.

At the physical point we get
 \be
    f_+(0)|_{SU(3)}^{PQ} = 0.9599(61)(32) ~ ,
    \label{eq:fplusPQ_SU3}
 \ee
where the first error is statistical and the second one is systematic coming from 
the uncertainties of the mass extrapolation and the $q^2$-dependence of the form 
factors.

We now discuss the analysis based on SU(2) ChPT.
First we note that Eq.~(\ref{eq:SU2}) holds for full QCD \cite{SU2} as well as for 
the PQ theory with $N_f = 2$.
In the latter case it can be verified by expanding $f_2^{PQ}$ [see Eq.~(\ref{eq:f2PQ})] 
in powers of $M_\pi^2 / M_K^2$. 
Thus we consider a SU(2) fit of the form (\ref{eq:SU2}) treating $F_+$ and $c_+$ 
as fitting parameters, and we apply it to our data with $M_\pi \lesssim 
0.4~\gev$.
Alternatively we add to Eq.~(\ref{eq:SU2}) a NNLO correction proportional to $M_\pi^4$ 
and apply the new fit to all lattice points.
The results are shown in Fig.~\ref{fig:fplus}(b).
It can be seen that the impact of the SU(2) NNLO correction is quite small up to 
$M_\pi \approx 0.5~\gev$ at variance with the corresponding SU(3) result shown 
in Fig.~\ref{fig:fplus}(a).
This finding signals a better convergence of SU(2) ChPT with respect to SU(3) for 
$f_+(0)$.

At the physical point we get 
 \be
    f_+(0)|_{SU(2)}^{PQ} = 0.9563(53)(13) ~ . 
    \label{eq:fplusPQ_SU2}
 \ee

The application of SU(2) and SU(3) ChPT yields results for $f_+(0)$, Eqs.~(\ref{eq:fplusPQ_SU3}) 
and (\ref{eq:fplusPQ_SU2}), which are consistent within the uncertainties.
By averaging the two results and adding the systematic errors in quadrature we get
 \be
    f_+^{PQ}(0) = 0.9581 \pm 0.0057_{\mbox{stat.}} \pm 0.0035_{\mbox{syst.}} ~ .
    \label{eq:fplus_PQ}
 \ee

We now present our estimates of the remaining sources of systematic effects.

{\it Finite Size.} We have performed a simulation at $M_\pi = 300~\mev$ using the 
volume $24^3 \cdot 48$, which corresponds to $M_\pi L \simeq 3.2$. 
We get $f_+(0) = 0.98633(362)$ using the pole-dominance fit (\ref{eq:pole}) and 
$f_+(0) = 0.98597(337)$ using the quadratic fit (\ref{eq:quadratic}).
We combine these values with the results shown in the second row of Table \ref{tab:fplus}, 
corresponding to the volume $32^3 \cdot 64$ with $M_\pi L \simeq 4.2$. 
Assuming a volume dependence of the form $A + B e^{-M_\pi L} / L^{3/2}$ we obtain
a residual FSE equal to $0.0018$, which we add (in quadrature) to the systematic 
error of Eq.~(\ref{eq:fplus_PQ}).

{\it Discretization.} We have performed a simulation at $M_\pi \simeq 470~\mev$ 
using a finer lattice spacing ($a \simeq 0.07~\mbox{fm}$).
We observe a systematic increase of the scalar form factor $f_0(q^2)$ at all values 
of $q^2$ and for all kaon masses.
In particular we get $f_+(0) = 0.99555(80)$ and $0.99518(95)$ using the pole-dominance 
(\ref{eq:pole}) and the quadratic (\ref{eq:quadratic}) fits, respectively.
We combine these values with the results shown in the fifth row of Table \ref{tab:fplus}. 
Assuming a linear fit in $a^2$ (which is consistent with the automatic ${\cal{O}}(a)$ 
improvement at maximal twist \cite{improvement}), we find a discretization effect 
equal to $0.0037$, which we add both to the central value and (in quadrature) 
to the systematic error of Eq.~(\ref{eq:fplus_PQ}).
Clearly a more detailed study of the scaling property of $f_+(0)$ would be beneficial 
in order to estimate better and to reduce further the discretization error.

{\it Quenching of the strange quark.} The effect of our PQ setup can be estimated 
within SU(3) ChPT, because, thanks to the AG theorem, the effect of quenching the 
strange quark is exactly known at NLO: at the physical point $f_2 - f_2^{PQ} = 
-0.0058$ ($\simeq 26\%$ of $f_2$).
This correction is added to the central value of Eq.~(\ref{eq:fplus_PQ}).
As for the ${\cal{O}}(p^6)$ term $\Delta f$, we have found evidence that the chiral 
logs, which are the most sensitive to quenching effects, are small compared to 
the contribution of the local terms (see Fig.~\ref{fig:fplus}(a)).
We estimate that the quenching effect on $\Delta f$ is at most $50\%$ of the same 
effect on $f_2$.
Thus we add (in quadrature) the value $0.0028$ ($\simeq 13\%$ of $\Delta f$) to the 
systematic error of Eq.~(\ref{eq:fplus_PQ}).
Note that this value is of the same size of the difference between our estimate 
of $\Delta f$ at $N_f = 2$ and the quenched one of Ref.~\cite{SPQCDR}.

Our final result is
 \be
    f_+(0) & = & 0.9560 \pm 0.0057_{\mbox{stat.}} \pm 0.0062_{\mbox{syst.}} \nn \\
           & = & 0.9560 \pm 0.0084 ~ ,
    \label{eq:final}
 \ee
which corresponds to $\Delta f = -0.0214(84)$.
Our determination agrees very well with the Leutwyler-Roos result \cite{LR} and 
with previous lattice calculations at $N_f = 0$ \cite{SPQCDR}, $N_f = 2$ 
\cite{JLQCD,RBC06,QCDSF} and $N_f = 2 + 1$ \cite{RBC08}.

Using the latest experimental determination of the product $|V_{us}| f_+(0) = 
0.21668(45)$ \cite{PDG,FlaviaNet} we get from (\ref{eq:final})
 \be
    |V_{us}| = 0.2267 \pm 0.0005_{\mbox{exp.}} \pm 0.0020_{f_+(0)} ~ .
    \label{eq:Vus}
 \ee
Combining this value with $|V_{ud}| = 0.97418(27)$ and $|V_{ub}| = 0.00393(36)$ 
from PDG2008 \cite{PDG} the CKM unitarity relation becomes
 \be
    |V_{ud}|^2 + |V_{us}|^2 + |V_{ub}|^2 = 1.0004 \pm 0.0015 .
    \label{eq:unitarity}
 \ee

\begin{figure}[!htb]

\vspace{0.25cm}

\centerline{\includegraphics[scale=0.45]{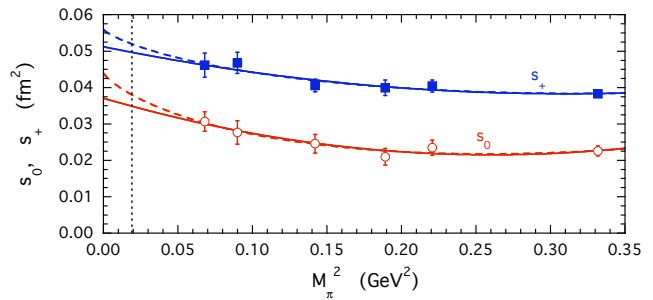}}

\caption{\it Results for the slopes $s_0$ (dots) and $s_+$ (squares) versus 
$M_\pi^2$ at $M_K = M_K^{ref}$.
The solid (dashed) line corresponds to the fit (\ref{eq:slopes}) with $d_{+,0} = 0$ 
($c_{+,0} = 0$).}

\label{fig:slopes}

\end{figure}

In conclusion we present our results for the slopes of the scalar ($s_0$) and vector 
($s_+$) form factors. 
Their light-quark mass dependence is illustrated in Fig.~\ref{fig:slopes} and it 
appears to be quite mild.
We have tried simple fitting functions of the form
 \be
    s_j = a_j + b_j M_\pi^2 + c_j M_\pi^4 + d_j M_\pi^2 ~ \mbox{log}(M_\pi^2) ~ ,
    \label{eq:slopes}
  \ee
where $a_j$, $b_j$, $c_j$ and $d_j$ are fitting parameters and $j = +, 0$.
The results of two fits, one with $d_{+,0} = 0$ and the other with $c_{+,0} = 0$, 
are shown in Fig.~\ref{fig:slopes}.

In terms of the dimensionless quantities $\lambda_{+,0} \equiv M_\pi^2 ~ s_{+,0}$ 
the extrapolation to the physical point and the evaluation of the systematic 
uncertainties yield 
 \be
    \lambda_0 & = & (12.8 \pm 2.2_{\mbox{stat.}} \pm 4.5_{\mbox{syst.}}) \cdot 10^{-3} ~ , \nn \\
    \lambda_+ & = & (23.7 \pm 2.3_{\mbox{stat.}} \pm 2.1_{\mbox{syst.}}) \cdot 10^{-3} ~ , 
    \label{eq:lambda}
 \ee
where the large systematic error on $\lambda_0$ is dominated by discretization effects.
Our results for both $\lambda_0$ and $\lambda_+$ agree very well with the latest 
experimental averages $\lambda_0^{exp.} = (13.4 \pm 1.2) \cdot 10^{-3}$ and 
$\lambda_+^{exp.} = (24.9 \pm 1.1) \cdot 10^{-3}$, obtained in 
Ref.~\cite{FlaviaNet} using data from KLOE, KTeV, ISTRA+ and 
NA48 experiments.

We thank all the ETMC members for fruitful discussions and the apeNEXT computer 
centres in Rome and Zeuthen for their invaluable technical help.
One of us (F.M.) also acknowledges the Consolider-Ingenio 2010 Program CPAN 
(CSD2007-00042).

\end{document}